\documentclass[prd,superscriptaddress,floatfix,preprintnumbers,altaffilletter,
twocolumn,nofootinbib]{revtex4}
\usepackage{graphicx,url,amssymb,amsmath,longtable,rotating,color,units,
wasysym,subfigure,epsfig,cancel,placeins}

\def\Omg{\Omega_\text{gw}}

\def\be{\begin{equation}}
\def\ee{\end{equation}}
\def\bi{\begin{itemize}}
\def\ei{\end{itemize}}
\def\ben{\begin{enumerate}}
\def\een{\end{enumerate}}
\def\i{\item{}}

\begin{document}

\title{Sensitivity curves for searches for gravitational-wave backgrounds}
\author{Eric Thrane}
\email{ethrane@ligo.caltech.edu}
\affiliation{LIGO Laboratory, California Institute of Technology, 
MS 100-36, Pasadena, California 91125, USA}

\author{Joseph D.\ Romano}
\email{joseph.romano@ligo.org}
\affiliation{Department of Physics and Astronomy and
Center for Gravitational-Wave Astronomy,
University of Texas at Brownsville, Texas 78520, USA}

\date{\today}

\begin{abstract}
We propose a graphical representation of detector sensitivity curves 
for stochastic gravitational-wave backgrounds that takes into account 
the increase in sensitivity that comes from {\em integrating over frequency} 
in addition to integrating over time.
This method is valid for backgrounds that have a power-law spectrum in the analysis band.
We call these graphs ``power-law integrated curves.''
For simplicity, we consider cross-correlation searches for unpolarized and 
isotropic stochastic backgrounds using two or more detectors.
We apply our method to construct power-law integrated sensitivity curves for 
second-generation ground-based detectors
such as Advanced LIGO, space-based detectors such as LISA and the Big Bang Observer, 
and timing residuals from a pulsar timing array.
The code used to produce these plots is available at \url{https://dcc.ligo.org/LIGO-P1300115/public}
for researchers interested in constructing similar sensitivity curves.
\end{abstract}

\maketitle

%%%%%%%%%%%%%%%%%%%%%%%%%%%%%%%%%%%%%%%%%%%%%%%%%%%%%%%%%%%%%%%%%%%%%%%%%%%%%%%
\section{Introduction}
\label{s:intro}

\begin{figure}[htbp]
    \psfig{file=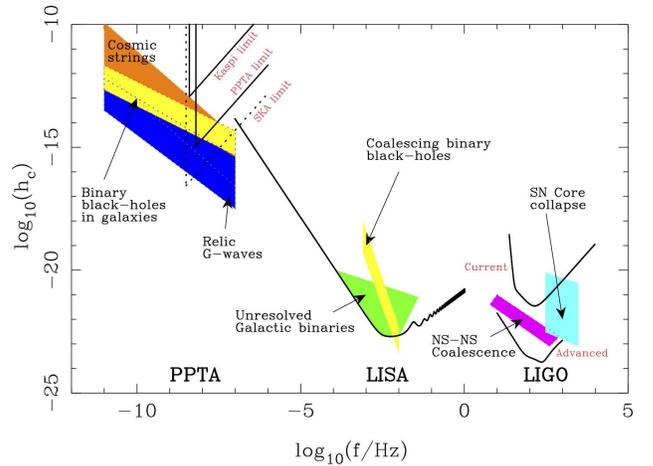, height=2.5in}
    \caption{
      Sensitivity curves for gravitational-wave observations and the 
      predicted spectra of various gravitational-wave sources, taken from \cite{hobbs}.
    }
    \label{f:sensitivity_curves}
\end{figure}

When discussing the feasibility of detecting gravitational waves 
using current or planned detectors, one often plots characteristic 
strain $h_c(f)$ curves of predicted signals (defined below in Eq.~\ref{e:hc}), and compares them to 
{\em sensitivity}  curves for different detectors.
The sensitivity curves are usually constructed by taking the ratio of 
the detector's noise power spectral density $P_n(f)$ 
to its sky- and polarization-averaged response 
to a gravitational wave ${\cal R}(f)$, defining 
$S_n(f)\equiv P_n(f)/{\cal R}(f)$ and an effective characteristic strain noise 
amplitude $h_n(f)\equiv \sqrt{f S_n(f)}$.
If the curve corresponding to a predicted signal $h_c(f)$ lies above the 
detector sensitivity curve $h_n(f)$ in some frequency band, then the signal 
has signal-to-noise ratio  $>$1.
An example of such a plot is shown in Fig.~\ref{f:sensitivity_curves}, 
which is taken from \cite{hobbs}.

\begin{figure}[htbp]
  \psfig{file=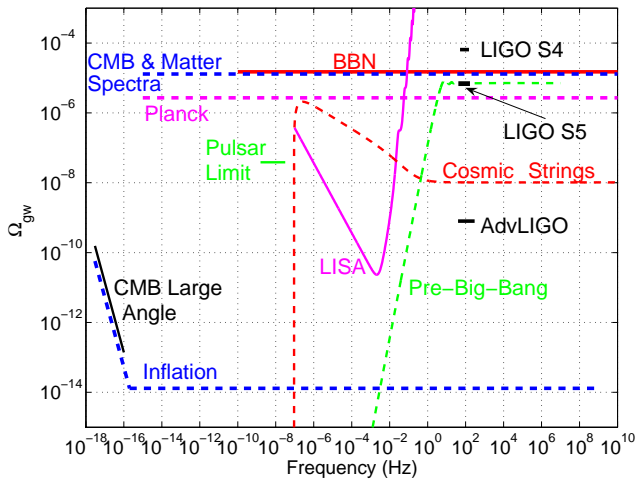, height=2.5in}
  \caption{
    Plot showing strengths of predicted gravitational-wave backgrounds 
    in terms of $\Omega_{\rm gw}(f)$ and the corresponding 
    sensitivity curves for different detectors, taken from \cite{stoch-S5}.
    Upper limits from various measurements, e.g., S5 LIGO Hanford-Livingston
    and pulsar timing, are shown as horizontal lines in the analysis band of each detector.
    The upper limits take into account integration over frequency, 
    but only for a single spectral index.
  }
  \label{f:landscape}
\end{figure}

For stochastic gravitational waves, which are typically searched
for by cross-correlating data from two or more detectors, one often 
adjusts the height of a sensitivity curve to take into account the 
total observation time (e.g., $T=1~{\rm yr}$ or $5~{\rm yr}$).
For uncorrelated detector noise, the expected (power) 
signal-to-noise ratio of a 
cross-correlation search for a gravitational-wave background for
frequencies between $f$ and $f+\delta{f}$ scales like $\sqrt{T\delta f}$.
So the effective characteristic strain noise amplitude $h_n(f)$ should be multiplied by a 
factor of $1/(T\delta f)^{1/4}$.
Also, instead of characteristic strain, one often plots the
predicted fractional energy density in gravitational waves 
$\Omega_{\rm gw}(f)$ as a function of frequency, 
which is proportional to $f^2 h^2_c(f)$ (see~Eq.~\ref{e:Omega-hc}).
An example of such a plot is shown in Fig.~\ref{f:landscape},
which is taken from \cite{stoch-S5}.

But for stochastic gravitational waves,
plots such as Figs.~\ref{f:sensitivity_curves} and \ref{f:landscape} do not always tell the full story.
Searches for gravitational-wave backgrounds also benefit from the {\em broadband} nature of the signal.
The integrated signal-to-noise ratio $\rho$ (see Eq.~\ref{e:snr}) 
also scales like $\sqrt{N_{\rm bins}}=\sqrt{\Delta{f}/\delta f}$, where $N_{\rm bins}$ is the 
number of frequency bins of width $\delta f$ in the total bandwidth $\Delta{f}$.
As we shall see below, the actual value of the proportionality constant depends on the 
spectral shape of the background and on the detector geometry (e.g., the separation and 
relative orientation of the detectors), in addition to the individual detector noise power spectral densities.
Since this improvement to the sensitivity is {\em signal dependent}, it is not always 
folded into the detector sensitivity curves,
even though the improvement in sensitivity can be significant.\footnote{To be clear, 
integration over frequency is always carried 
out in searches for stochastic gravitational-wave backgrounds, 
even though this is not always depicted in sensitivity curves.}
And when it is folded in, as in Fig~\ref{f:landscape}, a single spectral index is assumed, 
making it difficult to compare published limits with arbitrary models.
In other cases, limits are given as a function of spectral index, 
but the constrained quantity depends on an arbitrary reference frequency; see Eq.~\ref{e:OmegaPL}.

To illustrate the improvement in sensitivity that
comes from integrating over frequency,  consider 
the simple case of a white 
gravitational-wave background signal in white uncorrelated detector noise.
In this case, $\rho$ increases by precisely $\sqrt{N_{\rm bins}}$
compared to the single bin analysis.
For ground-based detectors like LIGO, typical values\footnote{The $\unit[0.25]{Hz}$ bin width typical of LIGO stochastic analyses is chosen to be sufficiently narrow that one can approximate the signal and noise as constant across the width of the bin, yet sufficiently wide that the noise can be approximated as stationary over the duration of the data segment.} of 
$\Delta f$ and $\delta f$ are  $\Delta f\!\approx\!100~{\rm Hz}$ 
and $\delta f\!\approx\!0.25~{\rm Hz}$, leading to $N_{\rm bins}\approx 400$,
and a corresponding improvement in $\rho$ of about $20$; see, e.g.,~\cite{stoch-S5}.
%A frequency bin width of $\unit[0.25]{Hz}$ is sufficiently narrow that one can approximate the signal and noise as constant across the width of the bin.
%At the same time, it is sufficiently wide that we can approximate the noise as stationary over the course of the data segment.
For colored spectra and non-trivial detector geometry the improvement 
will be less, but a factor of $\sim\!5$-10 increase in $\rho$
is not unrealistic.

\begin{figure}[htbp]
  \psfig{file=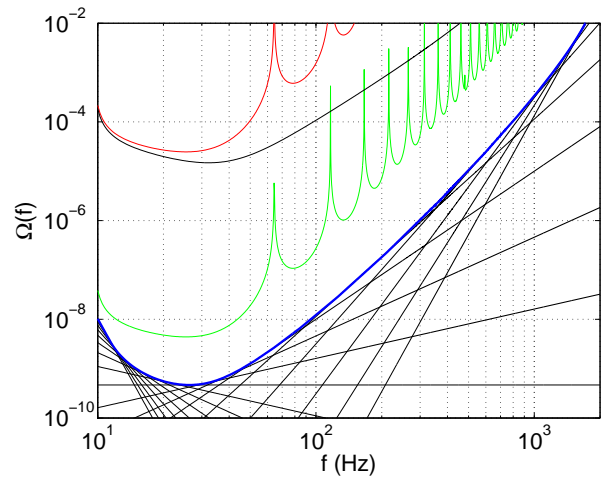, height=2.5in} 
  \caption{$\Omega_{\rm gw}(f)$
    sensitivity curves from different stages in a potential future 
    Advanced LIGO Hanford-LIGO Livingston correlation
    search for power-law gravitational-wave backgrounds.
    The top black curve is the single-detector sensitivity curve, 
    assumed to be the same for both H1 or L1.
    The red curve shows the sensitivity of the H1L1 detector pair
    to a gravitational-wave background, where the spikes are due to zeros in 
    the Hanford-Livingston overlap reduction function (see left panel, Fig.~\ref{f:LIGO-BBO}).
    The green curve shows the improvement in sensitivity that comes from integration 
    over an observation time of 1~year for a frequency bin size of $0.25$~Hz.
    The set of black lines are obtained by integrating over frequency for 
    different power law indices, assuming a signal-to-noise ratio $\rho =1$.
    Finally, the blue power-law integrated sensitivity curve is the envelope of the black lines.
    See Sec.~\ref{s:sensitivity}, Fig.~\ref{f:graphical_construction} for more details.}
  \label{f:preview_graphical_construction}
\end{figure}

In this paper, we propose a relatively simple way to graphically represent 
this improvement in sensitivity for gravitational-wave backgrounds that have 
a {\em power-law} frequency dependence in the sensitivity band of the detectors.
An example of such a ``power-law integrated sensitivity curve" is given 
in Fig.~\ref{f:preview_graphical_construction} 
for a correlation measurement 
between the Advanced LIGO detectors in Hanford, WA and Livingston, LA.
Details of the construction and interpretation of these curves will be
given in Sec~\ref{s:sensitivity}, Fig.~\ref{f:graphical_construction}.
We show this figure now for readers who might be anxious to get to the punchline.

In Sec.~\ref{s:formalism}, we briefly review the fundamentals of 
cross-correlation searches for gravitational-wave backgrounds, 
defining an effective strain noise power spectral density
$S_{\rm eff}(f)$ for a network of detectors.
For simplicity, we consider cross-correlation searches for 
unpolarized and isotropic stochastic backgrounds using 
two or more detectors.
In Sec.~\ref{s:sensitivity} we present a graphical method for 
constructing sensitivity curves for 
power-law backgrounds based on the expected signal-to-noise
ratio for the search, and we 
apply our method to construct new power-law integrated 
sensitivity curves for correlation measurements involving second-generation 
ground-based detectors such as Advanced LIGO, space-based detectors such as
the Big Bang Observer (BBO), 
and a pulsar timing array.
For completeness, we also construct a power-law integrated sensitivity 
curve for an autocorrelation measurement using LISA.
We conclude with a brief discussion in Sec.~\ref{s:discussion}.

%%%%%%%%%%%%%%%%%%%%%%%%%%%%%%%%%%%%%%%%%%%%%%%%%%%%%%%%%%%%%%%%%%%%%%%%%%%%%%%
\section{Formalism}
\label{s:formalism}

In this section, we summarize the fundamental properties
of a stochastic background and the correlated 
response of a network of detectors to such a background.
In order to keep track of the many different variables necessary 
for this discussion, we have 
included Table~\ref{tab:summary}, which summarizes key variables.

\begin{table*}
  \begin{tabular}{|c|l|}
  \hline
    {\bf variable} & {\bf definition} \\\hline
    $h_{ab}(t,\vec{x})$ & metric perturbation, Eq.~\ref{e:hab} \\\hline
    $h_A(f,\hat{k})$ & Fourier coefficients of metric perturbation, Eq.~\ref{e:hab} \\\hline
    $S_h(f)$ & strain power spectral density of a gravitational-wave background, Eq.~\ref{e:Sh} \\\hline
    $\Omg(f)$ & fractional energy density spectrum of a gravitational-wave background, Eq.~\ref{e:Omega(f)} \\\hline
    $h_c(f)$ & characteristic strain for gravitational waves, Eq.~\ref{e:hc} \\\hline
    $h(t)$ & detector response to gravitational waves, Eq.~\ref{e:h(t)} \\\hline
    $R_I^A(f,\hat{k})$ & detector response to a sinusoidal plane gravitational wave, Eq.~\ref{e:h(t)} \\\hline
    $\tilde{h}(f)$ & Fourier transform of $h(t)$, Eq.~\ref{e:h(f)} \\\hline
    $\Gamma_{IJ}(f)$ & overlap reduction function for the correlated response to a gravitational-wave background, 
    Eq.~\ref{e:Gamma} \\\hline
    ${\cal R}_I(f)$ & detector response to a gravitational wave averaged over polarizations and directions on the sky, 
    Eq.~\ref{e:R} \\\hline
    $P_{hI}(f)$ & detector power spectral density due to gravitational waves, Eq.~\ref{e:Ph} \\\hline
    $P_{nI}(f)$ & detector power spectral density due to noise, Eq.~\ref{e:snr} \\\hline
    $S_{\rm eff}(f)$ & effective strain noise power spectral density for a detector network, Eq.~\ref{e:Seff} \\\hline
    $h_{\rm eff}(f)$ & effective characteristic strain noise amplitude for a detector network, Eq.~\ref{e:heff} \\\hline
    $S_n(f)$ & strain noise power spectral density for a single detector, Eq.~\ref{e:Sn} \\\hline
    $h_n(f)$ & characteristic strain noise amplitude for a single detector, $h_n(f)\equiv \sqrt{fS_n(f)}$\\\hline
\end{tabular}
\caption{Summary of select variables with references to key equations.}
\label{tab:summary}
\end{table*}

\subsection{Statistical properties}
\label{s:characterizing}

In transverse-traceless coordinates, the metric perturbations $h_{ab}(t,\vec x)$ 
corresponding to a gravitational-wave background 
can be written as a linear superposition of 
sinusoidal plane gravitational waves with frequency $f$, 
propagation direction $\hat k$, and polarization $A$:
\begin{equation}
  \begin{split}
    &
    h_{ab}(t,\vec x)
    = \\
    &
    \int_{-\infty}^\infty df
    \int_{S^2} d^2\Omega_{\hat k}
    \sum_A
    h_A(f,\hat k)
    e^A_{ab}(\hat k)\,
    e^{i 2\pi f(t-\hat k\cdot \vec x/c)}\,,
  \end{split}
  \label{e:hab}
\end{equation}
where $e^A_{ab}(\hat k)$ are the gravitational-wave polarization tensors 
and $A=+,\times$ (see e.g.,~\cite{Allen-Romano:1999}).
The Fourier components $h_A(f,\hat k)$ are {\em random} fields whose 
expectation values define the statistical properties of the background.
Without loss of generality we can assume $\langle h_A(f,\hat k)\rangle=0$.
For unpolarized and isotropic stochastic backgrounds, 
the quadratic expectation values have the form
\begin{align}
&\langle h_A(f,\hat k) h_{A'}^*(f',\hat k')\rangle = 
\nonumber\\
&\qquad\qquad\qquad
\frac{1}{16\pi} 
\delta(f-f') \delta_{AA'} \delta^2(\hat k,\hat k') S_h(f) ,
\label{e:isotropicexpectationvalues}
\end{align}
where
\be
S_h(f) = \frac{3 H_0^2}{2\pi^2} \frac{\Omg(f)}{f^3} 
\label{e:Sh}
\ee
is the gravitational-wave power spectral density, and
\be
\Omg(f)
=\frac{1}{\rho_c}\frac{d\rho_{\rm gw}}{d\ln f} 
\label{e:Omega(f)}
\ee
is the fractional contribution of the energy density in 
gravitational waves to the total energy density needed to 
close the universe \cite{Allen-Romano:1999}.
(Throughout this paper we utilize single-sided power spectra.)
The variable $\rho_c$ denotes the critical energy density of 
the universe while $d\rho_\text{gw}$ denotes the energy density between $f$ and $f+df$.
In terms of the characteristic strain defined by
\be
h_c(f) \equiv \sqrt{f S_h(f)}\,,
\label{e:hc}
\ee
it follows that
\be
\Omg(f) = \frac{2\pi^2}{3 H_0^2}f^2h_c^2(f)\,.
\label{e:Omega-hc}
\ee
%

%%%%%%%%%%%%%%%%%%%%%%%%%%%%%%%%%%%%%%%%%%
\subsection{Power-law backgrounds}

In this paper, we will restrict our attention
to gravitational-wave backgrounds that can
be described by power-law spectra:
\begin{equation}
  \Omg(f) = \Omega_\beta 
  \left(\frac{f}{f_{\rm ref}}\right)^\beta ,
  \label{e:OmegaPL}
\end{equation}
where $\beta$ is the spectral index and $f_{\rm ref}$ is a reference frequency, 
typically set to $\unit[1]{yr^{-1}}$ for pulsar-timing observations and $\unit[100]{Hz}$ for ground-based detectors.
The choice of $f_\text{ref}$, however, is arbitrary and 
does not affect the detectability of the signal.

It follows trivially that 
the characteristic strain also has a power-law
form:
\begin{eqnarray}
  h_c(f) 
  =A_\alpha\left(\frac{f}{f_{\rm ref}}\right)^\alpha ,
  \label{e:hcPL}
\end{eqnarray}
where the amplitude $A_\alpha$ and spectral index
$\alpha$ are related to $\Omega_\beta$ and $\beta$ via:
\be
  \Omega_\beta = \frac{2\pi^2}{3 H_0^2}
  f_{\rm ref}^2\,A_\alpha^2\,,
  %A_\alpha = \left( \Omega_\beta \, \frac{3 H_0^2}{2\pi^2 f_{\rm ref}^2} \right)^{1/2}\,,
  \quad
  \beta = 2\alpha+2\,.
  %\alpha = (\beta-2)/2
  \label{e:alpha-beta}
\ee

For inflationary backgrounds relevant for cosmology, it is often assumed that
\be
\Omg(f)={\rm const}\,,
\ee
for which $\beta=0$ and $\alpha=-1$.
For a background arising from binary 
coalescence,
\be
\Omg(f) \propto f^{2/3}\,,
\ee
for which $\beta = 2/3$ and $\alpha=-2/3$.
This power-law dependence is applicable to 
super-massive black-hole coalescences targeted by pulsar 
timing observations as well as compact binary coalescences 
relevant for ground-based and space-based detectors.

%%%%%%%%%%%%%%%%%%%%%%%%%%%%%%%%%%%%%%%%%%%%%%%%%%%%%%%%%%%%%%%%%%%%%%%%%%%%%%%
\subsection{Detector response}
\label{s:detresp}

The response $h(t)$ of a detector to a passing gravitational 
wave is the 
convolution of the metric perturbations $h_{ab}(t,\vec x)$ with the 
impulse response $R^{ab}(t,\vec x)$:
\be
  \begin{split}
    h(t) 
    &\equiv \int_{-\infty}^\infty d\tau\int d^3y\,
    R^{ab}(\tau,\vec y)
    h_{ab}(t-\tau,\vec x-\vec y) 
    \\
    &= \int_{-\infty}^\infty df\int d^2\Omega_{\hat k}\sum_A
    R^A(f,\hat k)
    h_A(f,\hat k)
    e^{i2\pi f (t-\hat k\cdot \vec x/c)}\,,
  \end{split}
  \label{e:h(t)}
\ee
where $\vec x$ is the location of the 
measurement at time $t$.
The function $R^A(f,\hat k)$ 
is the detector response to a sinusoidal plane-wave
with frequency $f$, propagation direction $\hat k$,
and polarization $A$.
In the frequency domain, we have
\begin{align}
\tilde{h}(f) 
= \int d^2\Omega_{\hat k}\sum_A
R^A(f,\hat k)
h_A(f,\hat k)
e^{-i2\pi f \hat k\cdot \vec x/c}\,.
\label{e:h(f)}
\end{align}
%

%%%%%%%%%%%%%%%%%%%%%%%%%%%%%%%%%%%
\subsection{Overlap reduction function}
\label{s:overlap}

Given two detectors, labeled by $I$ and $J$, 
the expectation value of the cross-correlation 
of the detector responses
$\tilde h_I(f)$ and $\tilde h_J(f)$ is
\begin{equation}
\langle\tilde h_I(f)\tilde h_J^*(f')
\rangle
=\frac{1}{2}\delta(f-f')\Gamma_{IJ}(f)S_h(f)\,,
\end{equation}
where 
\begin{align}
&\Gamma_{IJ}(f)
\equiv
\nonumber\\
&\quad
\frac{1}{8\pi}\int d^2\Omega_{\hat k}\,
\sum_A
R^A_I(f,\hat k)R^A_J{}^*(f,\hat k)
e^{-i2\pi f\hat k\cdot (\vec x_I-\vec x_J)/c}
\label{e:Gamma}
\end{align}
is the {\em overlap reduction function} 
(see e.g., \cite{Christensen:1992,Flanagan:1993} in the context of ground-based interferometers).
Note that $\Gamma_{IJ}(f)$ is the transfer function between gravitational-wave strain power 
$S_h(f)$ and detector response cross-power
$C_{IJ}(f) = \Gamma_{IJ}(f) S_h(f)$.
%it is a measure of the efficiency with which the detector pair converts 
%gravitational-wave strain power to instrument cross power.
It is often convenient to define a {\em normalized} overlap
reduction function $\gamma_{IJ}(f)$ such that for two identical,
co-located and co-aligned detectors, $\gamma_{IJ}(0)=1$.
For identical interferometers with opening angle between the arms $\delta$, 
\be
\gamma_{IJ}(f) =(5/\sin^2\delta)\,\Gamma_{IJ}(f)\,.
\ee

\begin{figure}[htbp]
  \begin{center}
    \psfig{file=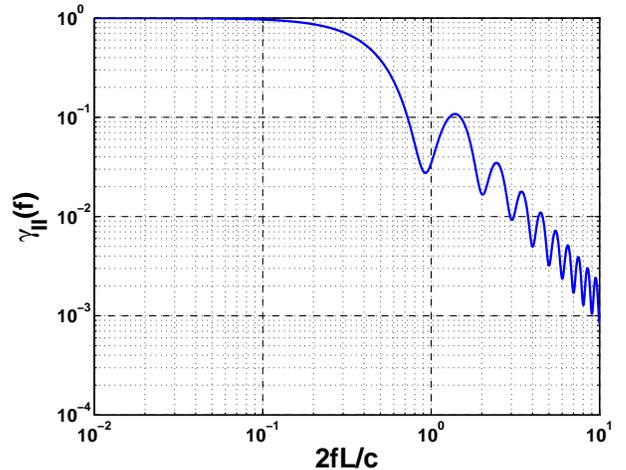, height=2.5in}
    \caption{
      A plot of the transfer function ${\cal R}_I(f)=\gamma_{II}(f)$ 
      normalized to unity for the strain response of an equal-arm Michelson interferometer.
      The dips in the transfer function occur around integer multiples of $c/(2L)$, 
      where $L$ is the arm length of the interferometer.
    }
    \label{f:gammaII}
  \end{center}
\end{figure}

For a single detector (i.e., $I=J$), we define
\be
\label{e:R}
  {\cal R}_I(f) \equiv \Gamma_{II}(f) ,
\ee
which is the transfer function between gravitational-wave strain 
power $S_h(f)$ and detector response auto power
\be
P_{hI}(f) = {\cal R}_I(f)S_h(f)\,.
\label{e:Ph}
\ee
Note that
${\cal R}_I(f)$ is the antenna pattern of 
detector $I$ averaged 
over polarizations and directions on the sky.
A plot of ${\cal R}_I(f)$ normalized to unity for the strain response of 
an equal-arm Michelson interferometer is shown in Fig.~\ref{f:gammaII}.

\begin{figure*}[htbp]
  \begin{tabular}{cc}
    \psfig{file=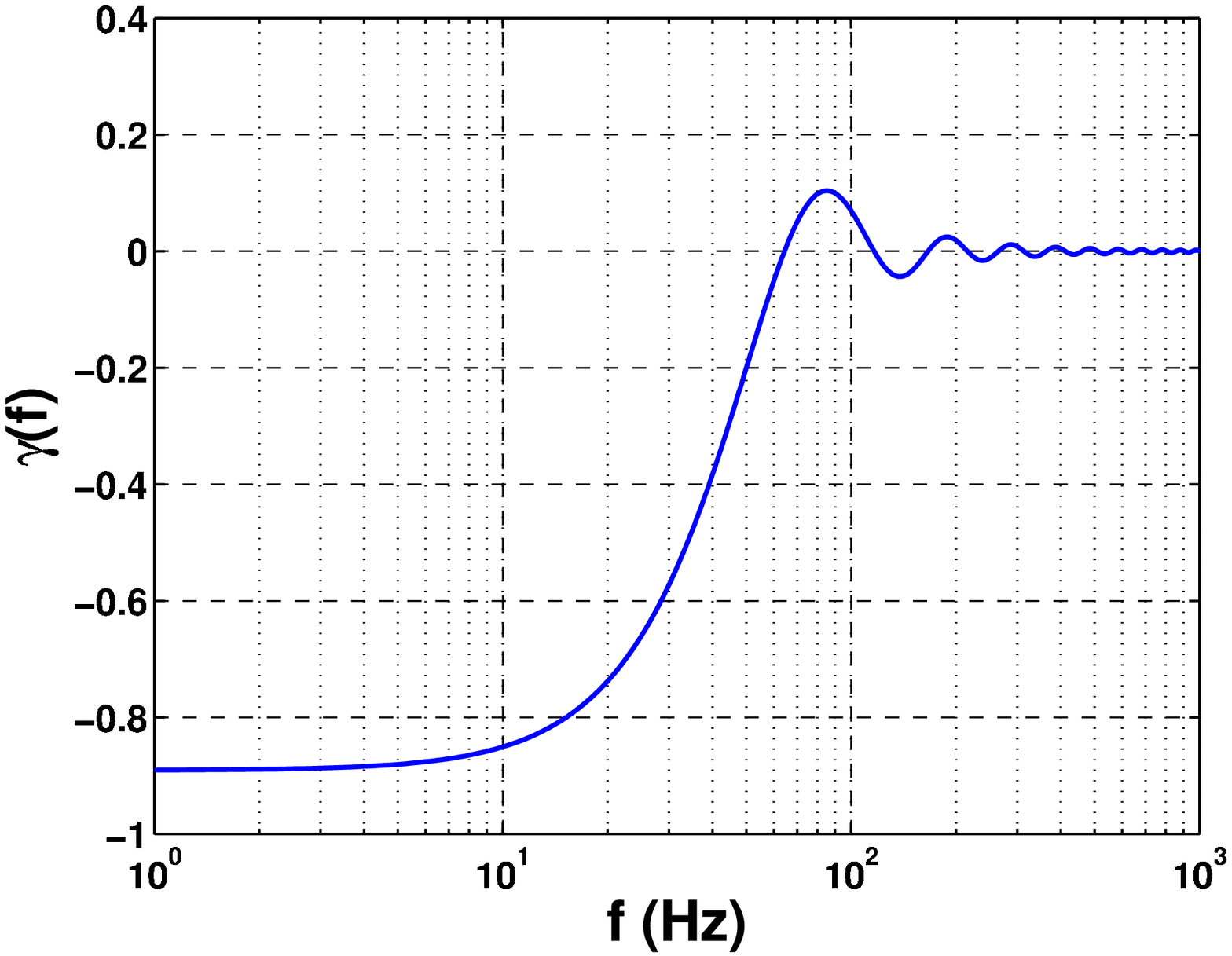, height=2.5in} &
    \psfig{file=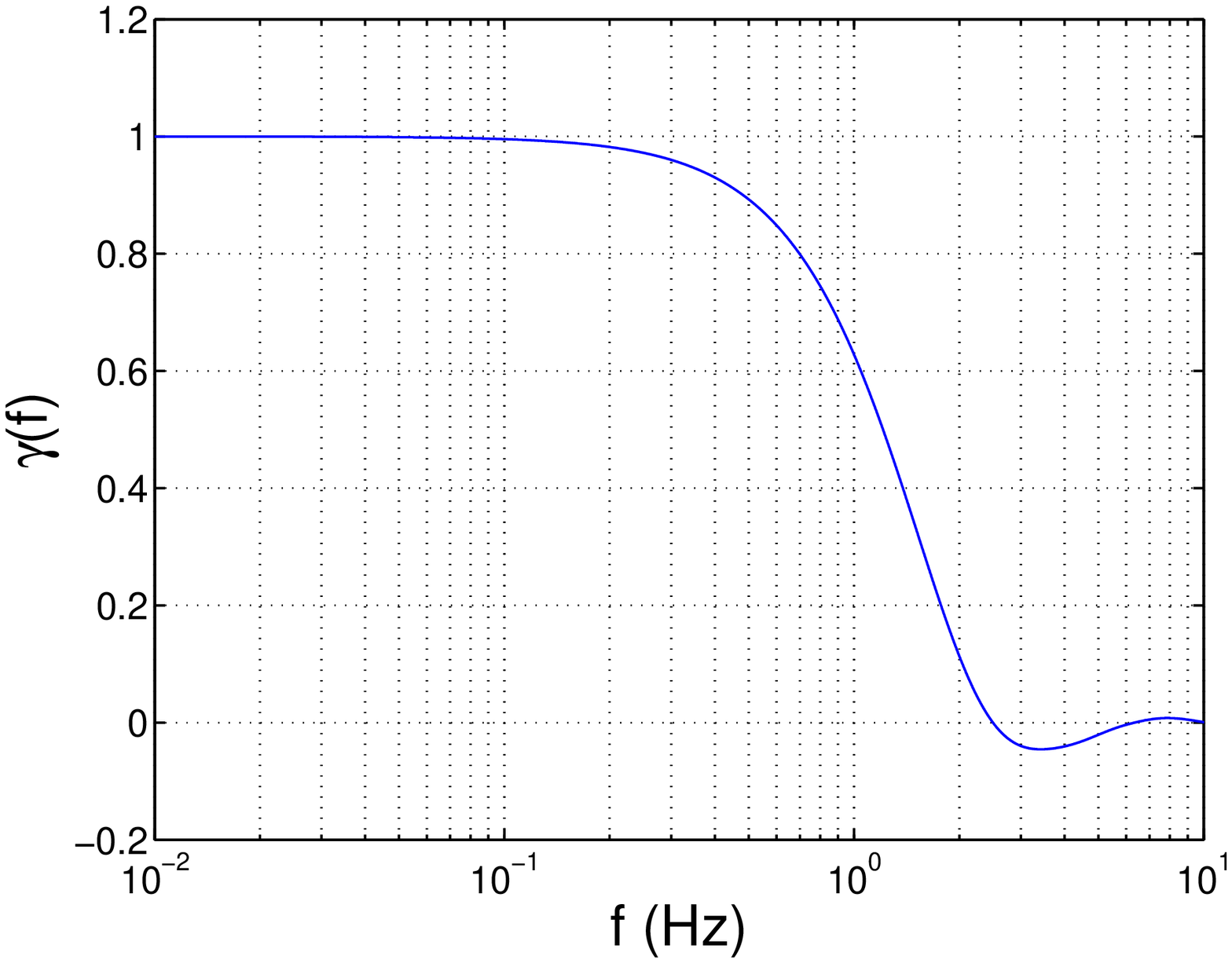, height=2.5in}\\
  \end{tabular}
  \caption{
    Left panel: Normalized overlap reduction function for the LIGO detectors 
    located in Hanford, WA and Livingston, LA.
    Right panel: Normalized overlap reduction function for two mini LISA-like
    Michelson interferometers located at opposite vertices of the BBO hexagram
    configuration.}
  \label{f:LIGO-BBO}
\end{figure*}

Detailed derivations and discussions of the overlap 
reduction functions 
for ground-based laser interferometers,
space-based laser interferometers, and 
pulsar timing arrays can be found 
in~\cite{Christensen:1992, Flanagan:1993, Allen-Romano:1999},
\cite{Cornish-Larson:2001, Finn-et-al:2009}, and 
\cite{Hellings-Downs:1983, Anholm-et-al:2009}, respectively.
In Fig.~\ref{f:LIGO-BBO} we plot the 
overlap reduction functions for
the strain response of the 
LIGO Hanford-LIGO Livingston detector pair in the 
long-wavelength limit (valid for frequencies below a few kHz)
and the strain response of a pair of 
mini LISA-like Michelson interferometers in the 
hexagram configuration of 
the Big Bang Observer (BBO), which is a proposed space-based
mission, whose goal is the direct 
detection of the cosmological gravitational-wave background 
\cite{Phinney-et-al:2004, Crowder-Cornish:2005, Cutler-Harms:2006}.
The two Michelson interferometers for the BBO overlap reduction
function 
are located at opposite vertices of a hexagram (`Star of David')
and have arm lengths $L=5\times 10^7~{\rm m}$ 
and opening angles $\delta = 60^\circ$.

\begin{figure*}[htbp]
  \begin{tabular}{cc}
      \psfig{file=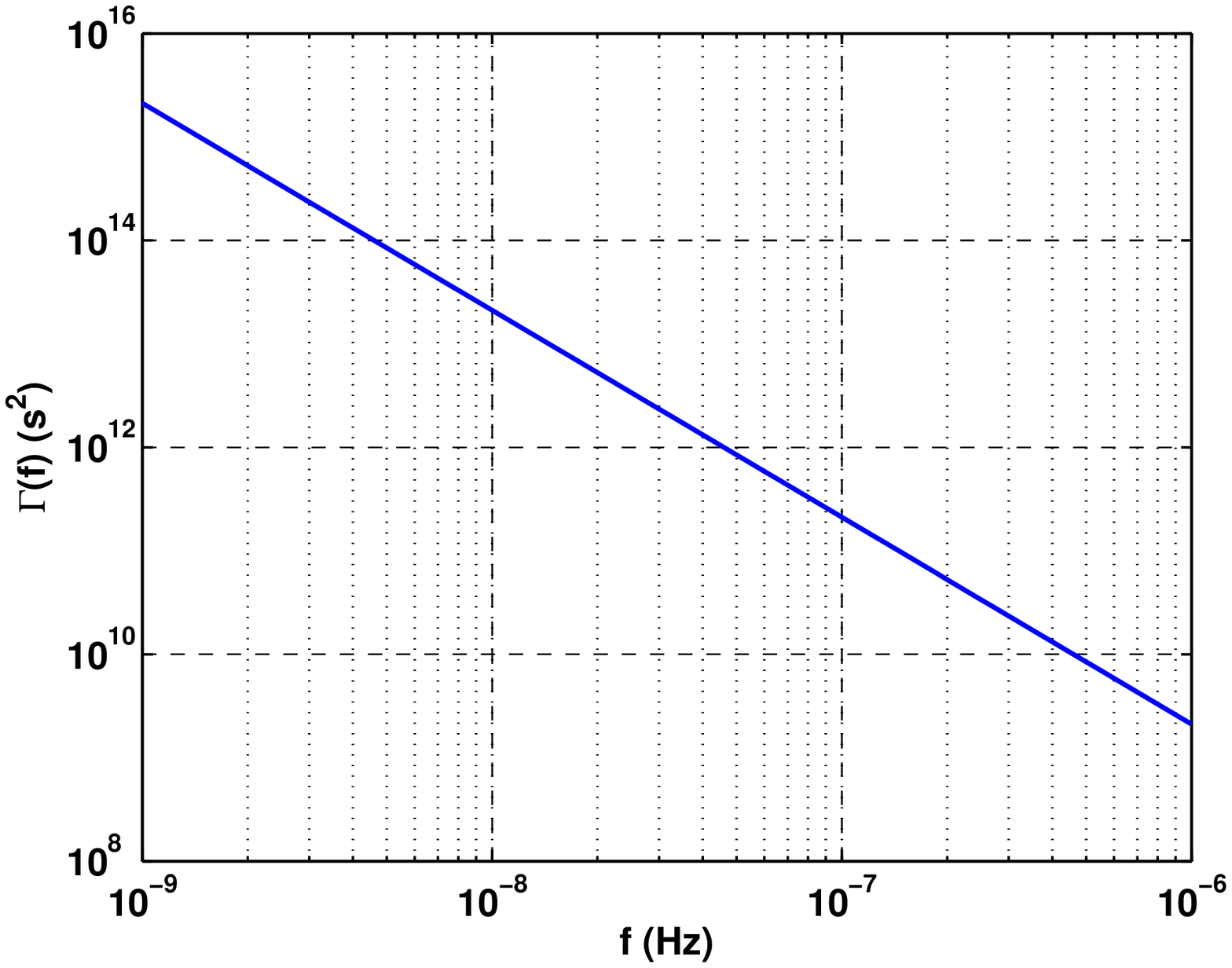, height=2.5in} &
      \psfig{file=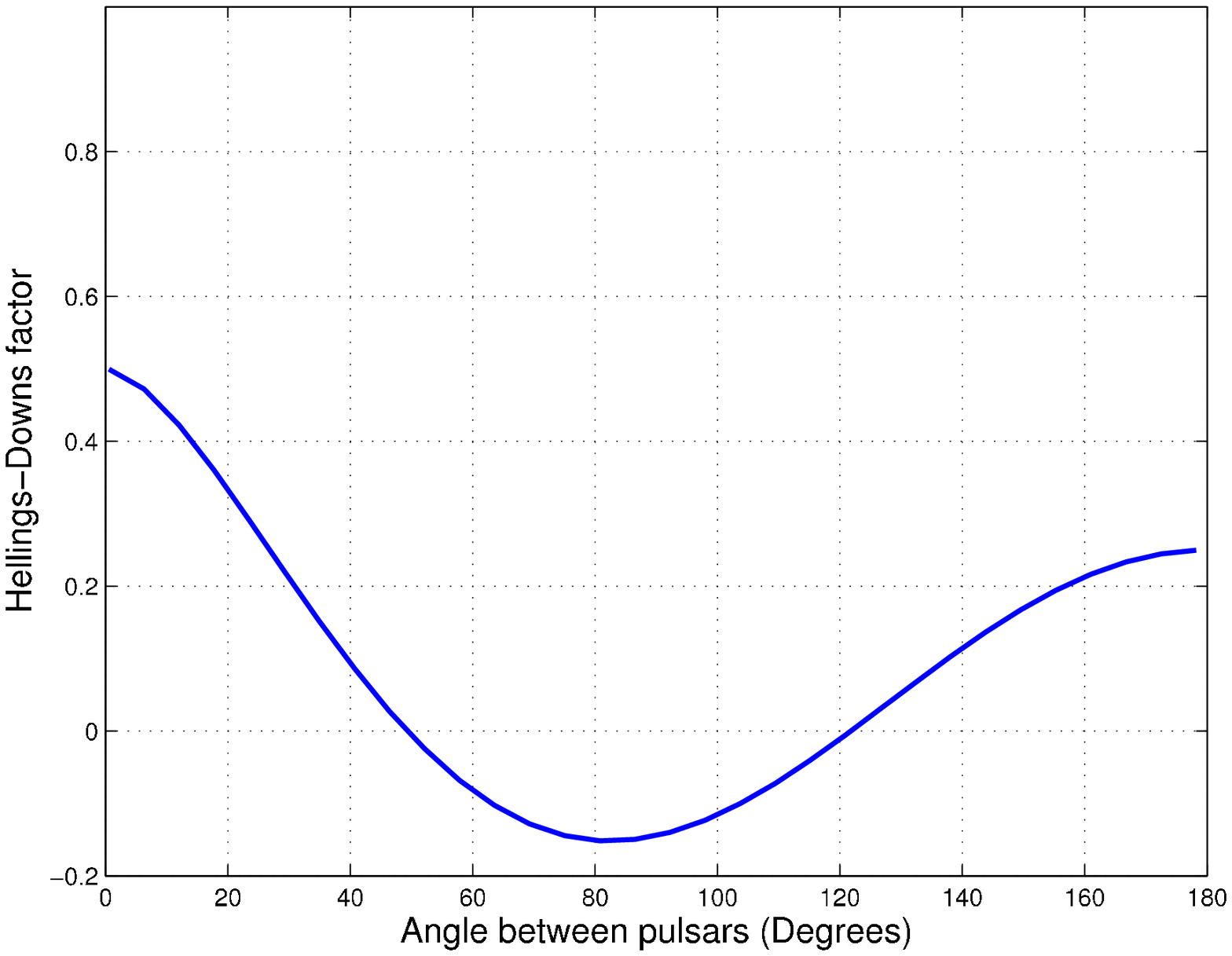, height=2.5in} 
  \end{tabular}
  \caption{
    Left panel: Overlap reduction function for a pair of pulsars,
    with $\zeta_{IJ}$ chosen to be 0.25. 
    Right panel: Hellings and Downs function $\zeta(\psi_{IJ})$.
    Note that the overlap reduction function is a function of frequency
    for a {\em fixed} pair of pulsars, while the Hellings and Downs function 
    is a function of the angle between two pulsars, and is {\em independent}
    of frequency.
  }
  \label{f:PTA-HD}
\end{figure*}

In Fig.~\ref{f:PTA-HD}, we plot both the overlap reduction function and 
the Hellings and Downs curve~\cite{Hellings-Downs:1983} for the 
timing response of a pair of pulsars in a pulsar timing array.
Assuming two pulsars are separated by an angle $\psi_{IJ}$ on the sky, then
to a very good approximation~\cite{Anholm-et-al:2009}:
\be
\Gamma_{IJ}(f)
=
\frac{1}{(2\pi f)^2}
\frac{1}{3}\,\zeta_{IJ}
\label{e:Gamma-PTA}
\ee
where
\be
\begin{split}
\zeta_{IJ}\equiv
&
\frac{3}{2}\left(\frac{1-\cos\psi_{IJ}}{2}\right)
\log
\left(\frac{1-\cos\psi_{IJ}}{2}\right)
\\
&\qquad\qquad
-\frac{1}{4}
\left(\frac{1-\cos\psi_{IJ}}{2}\right)
+\frac{1}{2}
+\frac{1}{2}\delta_{IJ}
\end{split}
\label{e:zetaIJ}
\ee
is the Hellings and Downs factor \cite{Hellings-Downs:1983}.
(The normalization is chosen so that for a single pulsar $\zeta_{II}=1$.)
%

%%%%%%%%%%%%%%%%%%%%%%%%%%%%%%%%%%%%%%%%%%%%%%%%%%%%%%%%%%%%%%%%%%%%%%%%%%%%%%%
\subsection{Signal-to-noise ratio}
\label{s:snr}

The expected (power) signal-to-noise ratio 
for a cross-correlation search for an unpolarized and 
isotropic stochastic background 
is given by \cite{Allen-Romano:1999}:
\begin{equation}
  \rho
  =\sqrt{2T}\left[
    \int_{f_\text{min}}^{f_\text{max}} df \,
    \frac{\Gamma^2_{IJ}(f) S_h^2(f)}
    {P_{nI}(f) P_{nJ}(f)}
        \right]^{1/2}\,,
\label{e:snr}
\end{equation}
where $T$ is the total (coincident) observation
time and $P_{nI}(f)$, $P_{nJ}(f)$ are 
the auto power spectral densities for the noise in 
detectors $I$, $J$.
The limits of integration $[f_\text{min},f_\text{max}]$ define the bandwidth of the detector.
This is the total {\em broadband} 
signal-to-noise ratio,
integrated over both time and frequency.
It can be derived as the expected signal-to-noise
ratio of a filtered cross-correlation of the output 
of two detectors, where the filter 
function is chosen so as to maximize the 
signal-to-noise ratio of the 
cross-correlation.\footnote{The above expression 
for $\rho$ assumes that the gravitational-wave 
background is {\em weak} compared to the instrumental
noise in the sense that 
$P_{hI}(f) \ll P_{nI}(f)$ for all frequencies in
the bandwidth of the detectors.}
For a {\em network} of detectors,
this generalizes to
\begin{equation}
\rho =
\sqrt{2T}
\left[
    \int_{f_{\text{min}}}^{f_{\text{max}}} df\, 
    \sum_{I=1}^M \sum_{J>I}^M
    \frac{\Gamma^2_{IJ}(f) S_h^2(f)}
    {P_{nI}(f) P_{nJ}(f)}
\right]^{1/2}\,,
\label{e:snr-network}
\ee
where $M$ the number of individual detectors,
and we have assumed the same coincident 
observation time $T$ for each detector.

The above expression for $\rho$ suggests 
the following definition of an {\em effective}
strain noise power spectral density 
for the detector network
\be
S_{\rm eff}(f) \equiv
\left[
\sum_{I=1}^M\sum_{J>I}^M 
  \frac{\Gamma_{IJ}^2(f)}{P_{nI}(f) P_{nJ}(f)}
  \right]^{-1/2}\,,
\label{e:Seff}
\ee
with corresponding strain noise amplitude 
\be
h_{\rm eff}(f)\equiv \sqrt{f S_{\rm eff}(f)}\,.
\label{e:heff}
\ee
In terms of $S_{\rm eff}(f)$, we have
\be
\rho
=\sqrt{2T\delta f}\,\sqrt{N_{\rm bins}}\,
\bigg\langle \frac{S_h^2}{S_{\rm eff}^2}\bigg\rangle^{1/2}\,,
\label{e:rho-Seff}
\ee
where $\langle\ \rangle$ denotes an 
average\footnote{Explicitly, $\langle X\rangle \equiv
(1/\Delta f)\int_{f_{\rm min}}^{f_{\rm max}} X(f)\, df$.}
over the total bandwidth of the detectors, $\Delta f=N_{\rm bins}\,\delta f$.
For the case of $M$ identical, co-located and co-aligned
detectors, things simplify further.
First,
\be
S_{\rm eff}(f) = \sqrt{\frac{2}{M(M-1)}}\, S_n(f)\,,
\ee 
where
\be
S_n(f) \equiv P_n(f)/{\cal R}(f)
\label{e:Sn}
\ee
is the strain noise power spectral density in
a single detector.
Second,
\be
\rho
=\sqrt{T\delta f}\,\sqrt{N_{\rm bins}}\,\sqrt{M(M-1)}\,
\bigg\langle \frac{S_h^2}{S_n^2}\bigg\rangle^{1/2}\,.
\label{e:scalings}
\ee
Thus, we see that the expected signal-to-noise
ratio scales linearly with the number of detectors
for $M\gg1$, the square-root of the total 
observation time, and the square-root of the number of 
frequency bins.
Note that $\sqrt{T\delta f}\sqrt{N_{\rm bins}} = \sqrt{T\Delta f}$,
which is the total time-frequency volume of the measurement.

%%%%%%%%%%%%%%%%%%%%%%%%%%%%%%%%%%%%%%%%%%%%%%%%%%%%%%%%%%%%%%%%%%%%%%%%%%%%%%%
\section{Power-law integrated curves}
\label{s:sensitivity}

\subsection{Construction}

The sensitivity curves that we propose are based on 
Eq.~\ref{e:snr-network} for the expected
signal-to-noise ratio $\rho$, 
applied to gravitational-wave backgrounds with power-law 
spectra.
These ``power-law integrated sensitivity curves" 
include the improvement in sensitivity 
that comes from the broadband nature of the signal, via 
the integration over frequency.
The following construction is cast in terms of $\Omg(f)$, 
but we note that power-law integrated curves can also 
easily be constructed for $h_c(f)$ or $S_h(f)$ using 
Eqs.~\ref{e:Sh} and \ref{e:hc} to convert between the different
quantities.
\ben
\i Begin with the detector noise power spectral densities
$P_{nI}(f)$, $P_{nJ}(f)$, and the overlap reduction functions 
$\Gamma_{IJ}(f)$ for two or more detectors.
Using Eq.~\ref{e:Seff}, first calculate the effective strain
power spectral density $S_{\rm eff}(f)$, and then convert it 
to energy density units $\Omega_{\rm eff}(f)$ using Eq.~\ref{e:Sh}.

\i Assume an observation time $T$, typically between $1$ and
$\unit[10]{yr}$.

\i For a set of power-law indices e.g., $\beta = \{-8, -7, \cdots 7, 8\}$ 
and some choice of reference frequency $f_{\rm ref}$,
calculate the value of the amplitude $\Omega_\beta$ 
such that the integrated signal-to-noise ratio has some fixed value, e.g., $\rho=1$.
Explicitly,
\be
\Omega_\beta = \frac{\rho}{\sqrt{2T}}\left[
\int_{f_{\rm min}}^{f_{\rm max}} df\, \frac{(f/f_{\rm ref})^{2\beta}}{\Omega_{\rm eff}^2(f)}
\right]^{-1/2}\,,
\label{e:OmegaBetaInt}
\ee
%where $[f_{\rm min},f_{\rm max}]$ define the bandwidth $\Delta f$ of the detectors.
Note that the choice of $f_\text{ref}$ is arbitrary and will not affect the sensitivity curve.

\i For each pair of values for $\beta$ and $\Omega_\beta$, plot 
$\Omg(f)=\Omega_\beta(f/f_{\rm ref})^\beta$ versus $f$.

\i The envelope of the $\Omg(f)$ power-law curves is the 
power-law integrated sensitivity curve 
for a correlation measurement using two or more detectors.
Formally, the power-law integrated curve is given by:
\begin{equation}
  \Omega_\text{PI}(f) = \max_\beta \left[ 
    \Omega_\beta
  \left(\frac{f}{f_\text{ref}}\right)^\beta \right].
\end{equation}
\een
{\em Interpretation}:
Any line (on a log-log plot) that is tangent to the power-law integrated sensitivity curve 
corresponds to a gravitational-wave background power-law spectrum with an 
integrated signal-to-noise ratio $\rho=1$.
This means that if the curve for a predicted background lies {\em everywhere below} the sensitivity curve, 
then $\rho<1$ for such a background.
On the other hand, if the curve for a predicted power-law background with spectral index $\beta$ lies 
{\em somewhere above} the sensitivity curve, then it will be observed with an expected value of $\rho= \Omega_\beta^{\rm pred}/\Omega_\beta>1$.
Graphically, $\Omega_\beta^{\rm pred}$ is the value of the predicted power-law spectrum evaluated at
$f_{\rm ref}$, while $\Omega_\beta$ is the value of the same power-law spectrum that is tangent to the 
sensitivity curve, also evaluated at $f_{\rm ref}$.

%%%%%%%%%%%%%%%%%%%%%%%%%%%%
\subsection{Plots}
\label{s:plots}

The calculation of a power-law integrated sensitivity curve
is demonstrated in the 
left-hand panel of Fig.~\ref{f:graphical_construction} for the 
Hanford-Livingston (H1L1) pair of Advanced LIGO detectors.
% OLD LOCATION OF f:graphical_construction
%
\begin{figure*}[htbp]
  \begin{tabular}{cc}
    \psfig{file=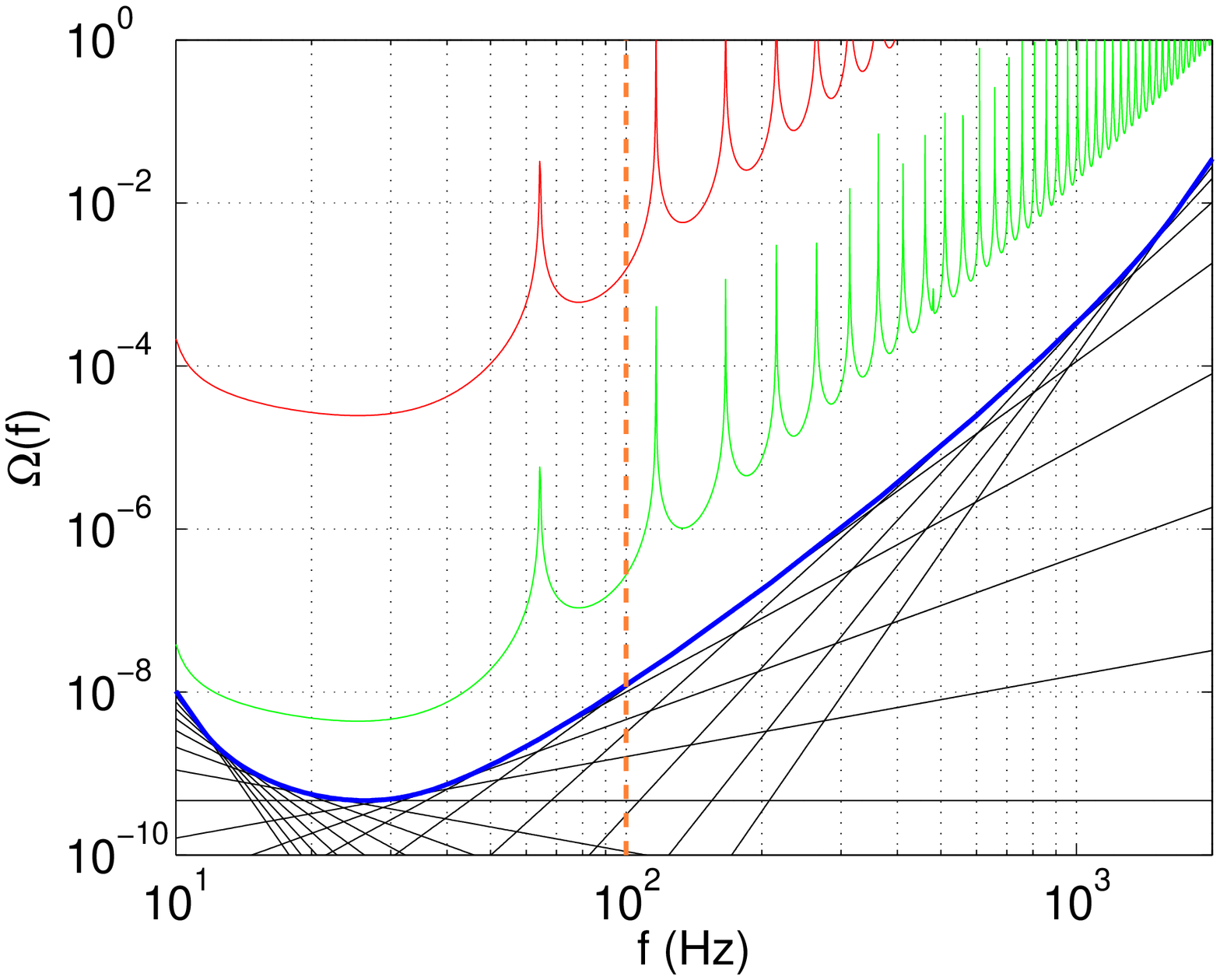, height=2.5in} &
    \psfig{file=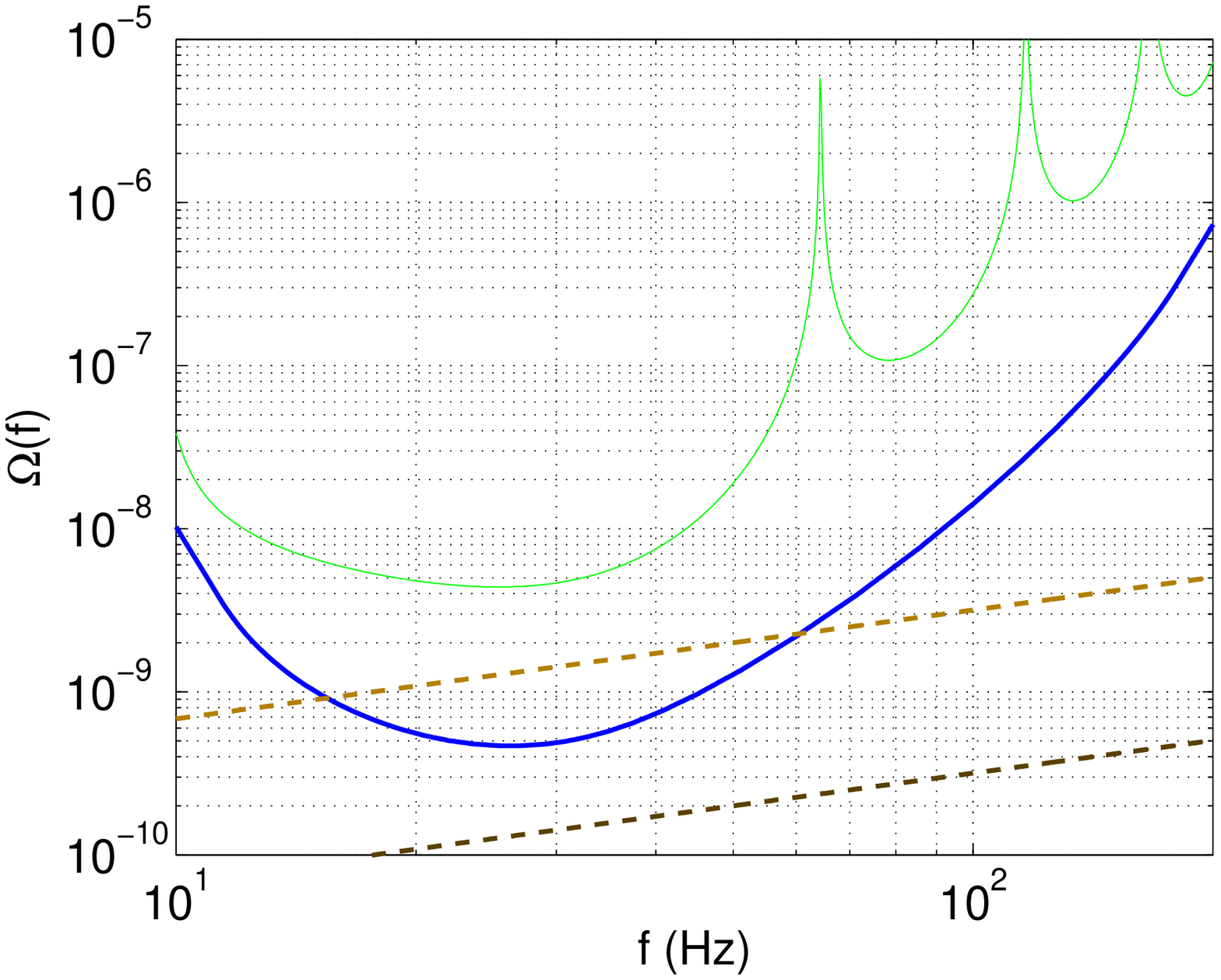, height=2.5in}
  \end{tabular}
  \caption{
    Left panel: $\Omg(f)$ sensitivity curves from different stages in a potential future Advanced LIGO H1L1 correlation
    search for power-law gravitational-wave backgrounds.
    The red line shows the effective strain spectral density $S_\text{eff}(f)=P_n(f)/|\Gamma_{\rm H1L1}(f)|$
    of the H1L1 detector pair to a gravitational-wave background signal converted to energy density
    $\Omega_{\rm eff}(f)$ via Eq.~\ref{e:Sh}.
    (The $P_n(f)$ used in this calculation is the design detector noise power spectral density for
    an Advanced LIGO detector, assumed to be the same for both H1 and L1.)
    The spikes in the red curve are due to zeroes in 
    the overlap reduction function $\Gamma_{\rm H1L1}(f)$, 
    which is shown in the left panel of Fig.~\ref{f:LIGO-BBO}.
    The green curve, $S_{\rm eff}(f)/\sqrt{2T\delta f}$, is obtained through the optimal combination of 
    one year's worth of data, assuming a frequency bin width of $\unit[0.25]{Hz}$ 
    as is typical~\cite{stoch-S5}.
    The vertical dashed orange line marks a typical Advanced LIGO reference frequency, $f_\text{ref}=\unit[100]{Hz}$.
    The set of black lines are obtained by performing the integration in 
    Eq.~\ref{e:OmegaBetaInt} for different power law indices $\beta$, requiring that $\rho =1$ to determine
    $\Omega_\beta$.
    Finally, the blue power-law integrated sensitivity curve is the envelope of the black lines.
    Right panel: a demonstration of how to interpret a power-law integrated curve.
    The thin green line and thick blue line are the same as in the left panel.
    The two dashed brown lines represent two different plausible signal models for gravitational-wave
    backgrounds arising from binary 
    neutron star coalescence; see, e.g.,~\cite{StochCBC}.
    In each case, $\Omg(f)\propto f^{2/3}$; however, the two curves differ by an order of magnitude 
    in the overall normalization of $\Omg(f)$.
    The louder signal will induce a signal-to-noise-ratio $\rho >1$ with an Advanced LIGO H1L1 correlation measurement 
    as it intersects the blue power-law 
    integrated curve---even though it falls below the time-integrated green curve.
    The weaker signal will induce a signal-to-noise-ratio $\rho <1$ with Advanced LIGO H1L1 as it is everywhere below the power-law
    integrated curve.
  }
  \label{f:graphical_construction}
\end{figure*}
Following steps $1$--$5$ above, we begin with the 
design detector noise power spectral density $P_n(f)$ 
for an Advanced LIGO detector \cite{aligo} 
(which we assume to be the same for both H1 and L1),
and divide by the absolute value of the 
H1L1 overlap reduction function to obtain 
the effective strain spectral density 
$S_{\rm eff}(f) = P_n(f)/|\Gamma_{\rm H1L1}(f)|$ of
the detector pair to a gravitational-wave background
(see Eq.~\ref{e:Seff}).
We then convert $S_{\rm eff}(f)$ to an energy density 
$\Omega_{\rm eff}(f)$ via Eq.~\ref{e:Sh}
to obtain the solid red curve.
After integrating $\unit[1]{yr}$ of coincident data, and assuming a frequency
bin width of $\unit[0.25]{Hz}$, 
we obtain the solid green curve, which is lower by a factor of $1/\sqrt{2T\delta f}$.
(The green curve, which depends on the somewhat arbitrary value of $\delta{f}$, can be thought of as an intermediate data product in LIGO analyses.)
Then assuming different spectral indices $\beta$,
we integrate over frequency (see~Eq.~\ref{e:OmegaBetaInt}),
setting $\rho=1$ to determine the amplitude $\Omega_\beta$ 
of a power-law background.
This gives us the set of black lines for each power law
index $\beta$.
The blue power-law integrated curve is the envelope of these black lines.

The right-hand panel of Fig.~\ref{f:graphical_construction} illustrates how to interpret 
a power-law integrated sensitivity curve.
We replot the green and blue curves from the left-hand panel, 
which respectively represent the time-integrated and power-law integrated sensitivity
of an Advanced LIGO H1L1 correlation measurement to a gravitational-wave background.
Additionally, we plot two theoretical spectra of the form $\Omg(f)\propto f^{2/3}$, 
which is expected for a background due to compact binary coalescences.
The dark brown line corresponds to a somewhat pessimistic scenario in which Advanced LIGO, 
running at design sensitivity, would detect $\approx10$ individual binary neutron star 
coalescences per year of science data~\cite{StochCBC}.
The light brown line represents a somewhat optimistic model in which Advanced LIGO, 
running at design sensitivity, would detect $\approx100$ individual binary neutron star 
coalescences per year of science data~\cite{StochCBC}.
(A binary-neutron-star detection rate of $\unit[40]{yr^{-1}}$ is considered a realistic 
rate for Advanced LIGO~\cite{cbc_rates}.)
The light-brown curve intersects the blue power-law integrated curve, 
indicating that the somewhat optimistic model will induce a signal-to-noise ratio $\rho>1$.
The dark brown curve falls below the blue power-law integrated curve, 
indicating that the somewhat pessimistic model will induce a signal-to-noise ratio $\rho<1$.
Note that {\em neither} curve intersects the green time-integrated sensitivity curve.

\medskip
In the following subsections, 
we plot power-law integrated sensitivity curves for several upcoming or proposed experiments:
networks of Advanced LIGO detectors (Fig.~\ref{fig:networks}),
BBO (Fig.~\ref{fig:plots}, top panel),
LISA (Fig.~\ref{fig:plots}, middle panel), and a network of pulsars from a pulsar timing array (Fig.~\ref{fig:plots}, bottom panel).

\subsubsection{Advanced LIGO networks}

For the Advanced LIGO networks, we use the design detector noise power spectral density 
$P_n(f)$ taken from~\cite{aligo} assumed to be the same for every detector in the network.
We consider three networks: 
H1L1 (just the US aLIGO detectors), 
H1H2 (a hypothetical co-located pair of aLIGO detectors), and 
H1L1V1K1 (the US aLIGO detectors plus detector pairs created with Virgo V1 and 
KAGRA K1).\footnote{We have taken the location and orientation of the KAGRA
detector to be that of the TAMA~300-m interferometer in Tokyo, Japan.
We have not included the planned LIGO India detector~\cite{india} in this 
network, as the precise LIGO-India site has not yet been decided upon.}
In reality, Virgo and KAGRA are expected to have different noise curves than aLIGO, 
but we assume the same aLIGO noise for each detector in order to show how the sensitivity 
curve changes by adding additional identical detectors to the network.
Given this assumption, the effective strain power spectral density 
can be written as 
\be
S_{\rm eff}(f) = {P_n(f)}/{{\cal R}_{\rm eff}(f)}\,,
\ee
where
\be
{\cal R}_{\rm eff}(f) = 
\left[\sum_{I=1}^M \sum_{J>I}^M \Gamma_{IJ}^2(f)\right]^{1/2}
\ee
is the sky- and polarization-averaged response of 
the network to a gravitational-wave background.
A plot of the various overlap reduction functions $\gamma_{IJ}(f)$
and ${\cal R}_{\rm eff}(f)$ for the H1L1V1K1 network 
are given in Fig.~\ref{f:network_orf}.
\begin{figure*}[hbtp!]
  \begin{tabular}{cc}
    \psfig{file=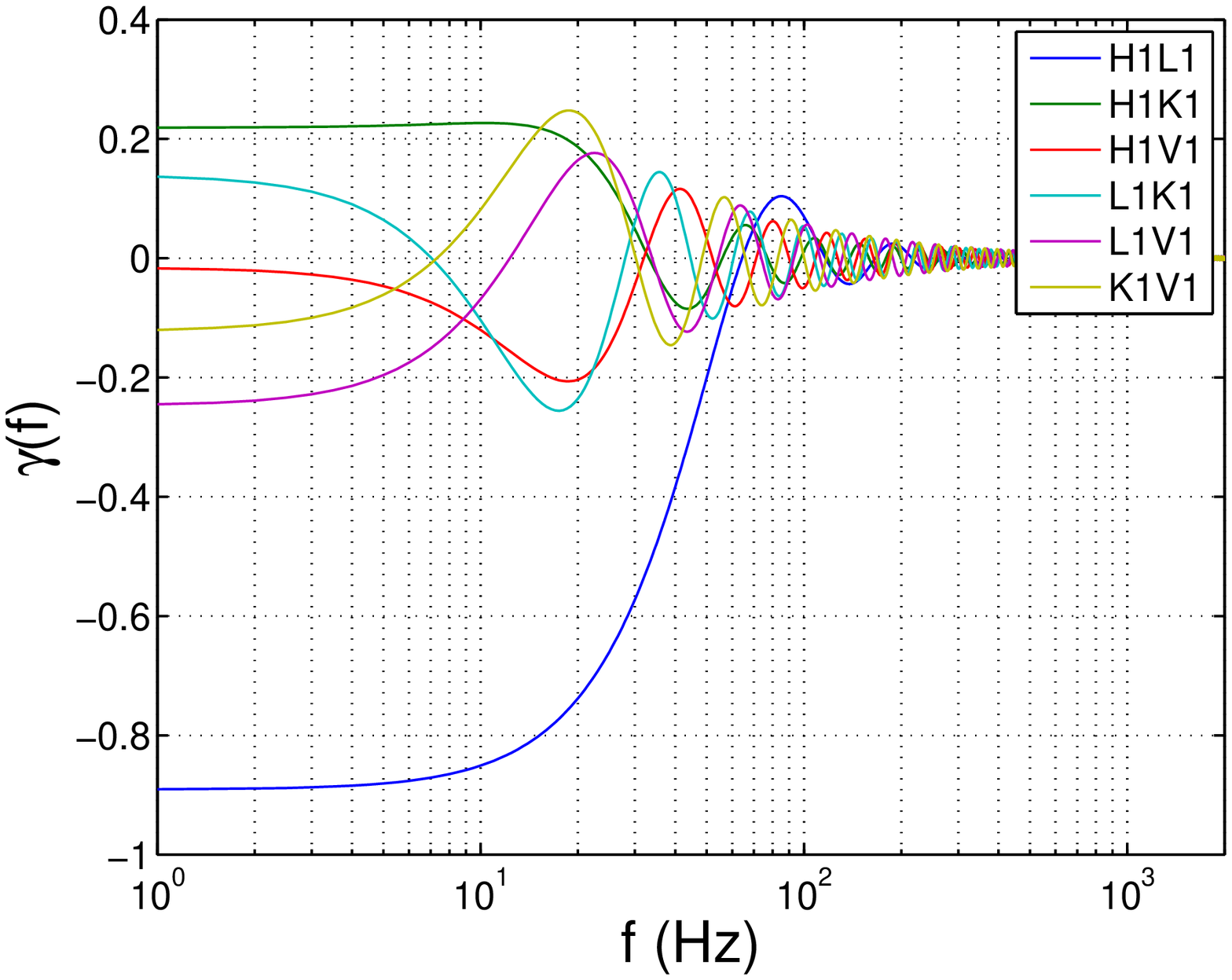, height=2.5in} &
    \psfig{file=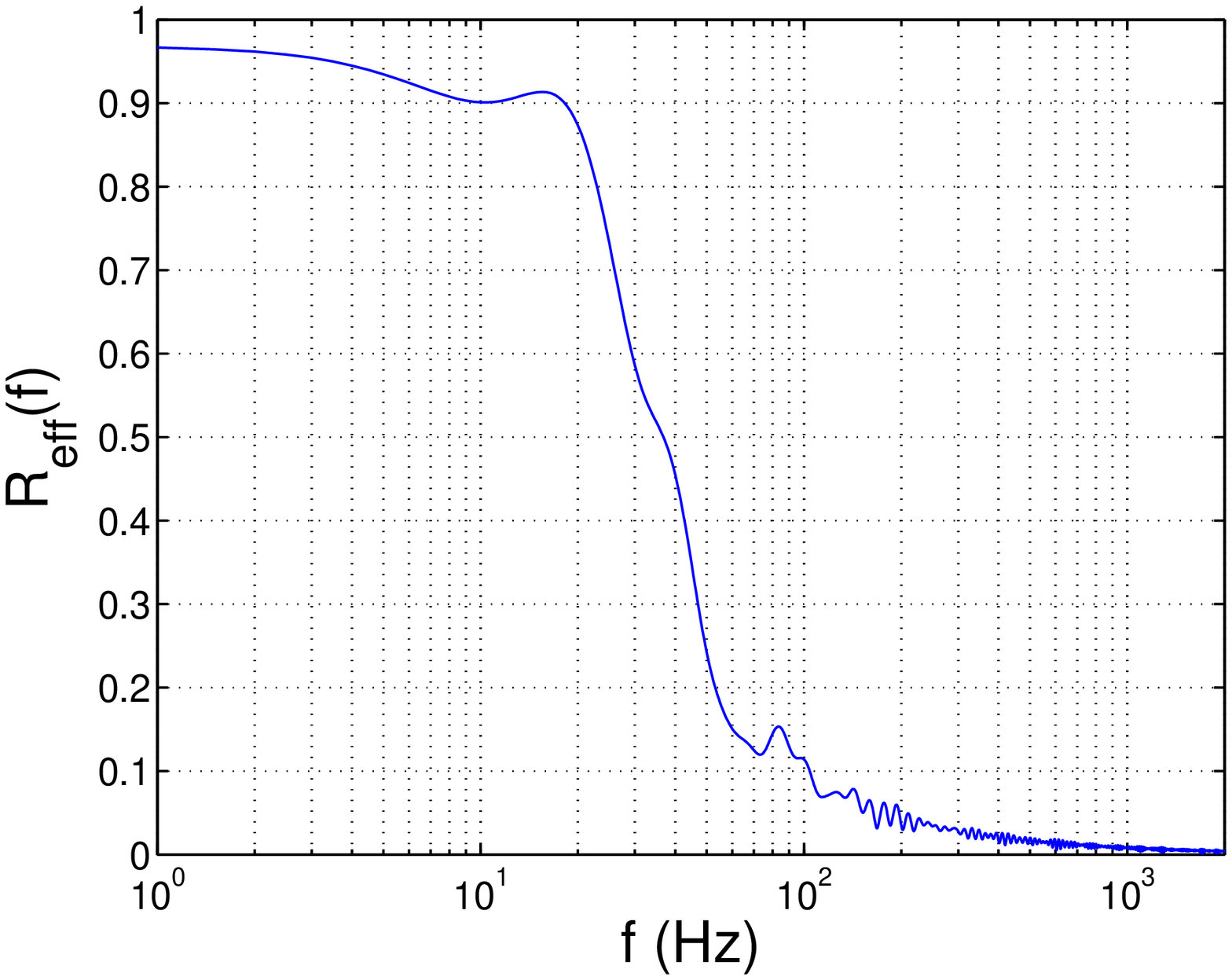, height=2.5in}
  \end{tabular}
  \caption{
    Left panel: Individual normalized overlap reduction functions for the six 
    different detector pairs comprising the H1L1K1V1 network.
    Right panel: Sky- and polarization-averaged response 
    of the H1L1V1K1 network to a gravitational-wave background.
  }
  \label{f:network_orf}
\end{figure*}
The resulting power-law integrated sensitivity curves are shown 
in Fig.~\ref{fig:networks}.
\begin{figure}[htbp]
  \psfig{file=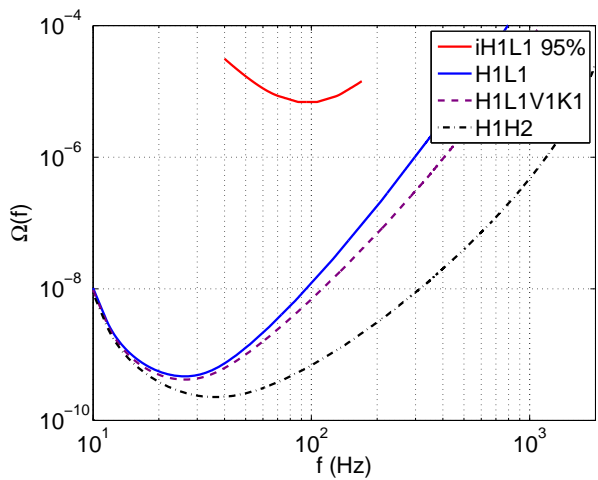, height=2.5in}
  \caption{
    Different networks of advanced detectors assuming $T=\unit[1]{yr}$ of observation.
    We also include 95\% CL limits from initial LIGO for comparison~\cite{stoch-S5}.
  }
  \label{fig:networks}
\end{figure}

\subsubsection{Big Bang Observer (BBO)}

For the BBO sensitivity curve, the noise power spectral density
for the two Michelson interferometers is taken to be
\be
P_n(f) =
\frac{4}{L^2}\left[
(\widetilde{\delta x})^2
+\frac{(\widetilde{\delta a})^2}{(2\pi f)^4}
\right]\,,
\label{e:BBO-noise}
\ee
where 
\begin{align}
(\widetilde{\delta x})^2 
&= 2\times 10^{-34}~\frac{{\rm m}^2}{{\rm Hz}}\,,
\\
(\widetilde{\delta a})^2 
&= 9\times 10^{-34}~\frac{{\rm m}^2}{{\rm s}^4\cdot {\rm Hz}}
\end{align}
are the position and acceleration noise
(see Table~II from \cite{Crowder-Cornish:2005}) and
$L=5\times 10^7~{\rm m}$ is the arm length.
Following \cite{Cutler-Harms:2006}, we have included an 
extra factor of 4 multiplying the first term 
in Eq.~\ref{e:BBO-noise}, which corresponds to 
high-frequency noise 4 times larger than shot noise alone.
The overlap reduction function for the Michelson interferometers
located at opposite vertices of the BBO hexagram
is shown in the right panel of Fig.~\ref{f:LIGO-BBO}.
The power-law integrated curve for BBO is given in Fig.~\ref{fig:plots},
top panel.

\subsubsection{LISA}

For LISA, the analysis is necessarily different since the standard 
cross-correlation technique used for multiple detectors such as 
an Advanced LIGO network, BBO, or a pulsar timing array 
is {\em not} possible for a single LISA constellation.
This is because the two independent Michelson interferometers that one
can synthesize from the six links of the standard equilateral LISA 
configuration are rotated at $45^\circ$ with respect one another, 
leading to zero 
cross-correlation for an isotropic gravitational-wave background for
frequencies below about $c/2L = 3\times 10^{-2}~{\rm Hz}$ \cite{Cutler:1998}.
It is possible, however, to construct a combination of the LISA 
data whose response to gravitational waves is {\em highly suppressed}
at these frequencies, and hence can be used as a real-time
{\em noise monitor} for LISA \cite{Tinto-Armstrong-Estabrook:2000, Hogan-Bender:2001}.
It is also possible to exploit the differences between the transfer 
function and spectral shape of a gravitational-wave background 
and that due to instrumental noise and/or an astrophysical foreground 
(e.g., from galactic white-dwarf binaries) to discriminate a
gravitational-wave background from these other noise 
contributions~\cite{Adams-Cornish:2010, Adams-Cornish:2013}.

For the ideal case of an autocorrelation measurement in a single
detector assuming perfect subtraction of instrumental noise
and/or any unwanted astrophysical foreground,
Eq.~\ref{e:snr} for the expected signal-to-noise ratio is replaced by
\begin{equation}
  \rho
  =\sqrt{T}\left[
    \int_0^\infty df\, 
    \frac{{\cal R}^2(f) S_h^2(f)}
    {P_{n}^2(f)}
        \right]^{1/2}\,,
\label{e:snr-LISA}
\end{equation}
where ${\cal R}(f)\equiv\Gamma(f)$ is the transfer function of the
detector and $P_n(f)$ is its noise power spectral density.
(The $\sqrt{2}$ reduction in $\rho$ compared to a
cross-correlation analysis is due to the use of data from
only one detector instead of two.)
For standard LISA,
\be
P_n(f) =
\frac{1}{L^2}\left[
(\widetilde{\delta x})^2
+\frac{4 (\widetilde{\delta a})^2}{(2\pi f)^4}
\right]\,,
\label{e:LISA-noise}
\ee
where 
\begin{align}
(\widetilde{\delta x})^2 
&= 4\times 10^{-22}~\frac{{\rm m}^2}{{\rm Hz}}\,,
\\
(\widetilde{\delta a})^2 
&= 9\times 10^{-30}~\frac{{\rm m}^2}{{\rm s}^4\cdot {\rm Hz}}
\end{align}
are the position and acceleration 
noise~\cite{Cornish-Larson:2001, Crowder-Cornish:2005} 
and $L=5\times 10^9~{\rm m}$ is the arm length.
The transfer function ${\cal R}(f)$ is taken from 
Fig.~\ref{f:gammaII} restricted to the LISA band,
$10^{-4}~{\rm Hz}<f< 10^{-1}~{\rm Hz}$.
Using the above expression for $\rho$ and following 
the same steps from the previous subsection for the 
construction of a power-law integrated curve, we 
obtain the sensitivity curve for LISA given in 
Fig.~\ref{fig:plots}, middle panel.

Note that the minimum value of $\Omega(f)$ shown in 
this plot is about a factor of 10 times smaller than 
the value of $\Omg(f)\approx 2\times 10^{-13}$
reported in \cite{Adams-Cornish:2010, Adams-Cornish:2013}.
Part of this difference is due to our use of 
$\rho=1$ for the sensitivity curve, 
while their value of $\Omg(f)$ corresponds to 
a {\em strong} (several $\sigma$) detection having a Bayes factor $\ge 30$.
The remaining factor can probably be attributed to 
the marginalization over the instrumental noise 
and galactic foreground parameters in \cite{Adams-Cornish:2010, Adams-Cornish:2013},
while Eq.~\ref{e:snr-LISA} assumes that we know these
parameters perfectly.

\subsubsection{Pulsar timing array}

For the pulsar timing array sensitivity curve, 
we consider a network of 20~pulsars 
taken from the International Pulsar Timing Network (IPTA)~\cite{Hobbs-et-al:2010}, 
which we assume have identical white timing noise power spectral densities,
\be
P_n(f) = 2\Delta t\,\sigma^2\,,
\ee
where $1/\Delta t$ is the cadence of the measurements, 
taken to be $20~{\rm yr}^{-1}$, and $\sigma$ is the root-mean-square 
timing noise, taken to be $100~{\rm ns}$.
We note that the pulsar timing network we envision may be 
somewhat optimistic as $\unit[100]{ns}$ root-mean-square timing noise is ambitious.
Also, we do not include the effects of fitting each pulsar's period $P$ and spin-down rate $\dot{P}$ to a timing model, which introduces both non-stationarity in the timing residuals and loss of sensitivity~\cite{haasteren_levin}.
Nevertheless, one can still write down an analogous expression to Eq.~\ref{e:snr-network} including these effects~\cite{siemens}.

Since the timing noise power spectral densities are 
identical, it follows that
\be
S_{\rm eff}(f) = {S_n(f)}
\,\left[\sum_{I=1}^M \sum_{J>I}^M \zeta_{IJ}^2\right]^{-1/2}\,,
\ee
where
\be
S_n(f) = {P_n(f)}/{{\cal R}(f)} = 
12\pi^2 f^2\, P_n(f)
\ee
% 
%\be
%h_{\rm eff}(f) = 
%\left(\sum_{I=1}^M \sum_{J>I}^M \zeta_{IJ}^2\right)^{-1/4}
%\sqrt{24\pi^2\Delta t}\,
%\sigma f^{3/2}\,,
%\ee
%
and $\zeta_{IJ}$ are the Hellings and Downs factors 
for each pair of pulsars in the array.
For our choice of 20 pulsars,
\be
\sum_{I=1}^M \sum_{J>I}^M \zeta_{IJ}^2
= 4.74\,,
\ee
which can thought of as the {\em effective} number of 
pulsar pairs for the network.
Finally, we assume a total observation time $T=5~{\rm yr}$, which sets the 
lower frequency limit of $S_{\rm eff}(f)$.
Given these parameters, we expect the pulsar timing array to be operating in the 
``intermediate signal limit''~\cite{siemens}.
We therefore utilize the scaling laws from Fig.~2 in Ref.~\cite{siemens} 
to adjust the power-law integrated curves, since Eqs.~\ref{e:snr}, \ref{e:snr-network}
for $\rho$ are valid in the {\em weak-signal limit} and 
overestimate the expected signal-to-noise
ratio by a factor of $\approx5$ for an observation of $T=5~{\rm yr}$.
The power-law integrated curve for IPTA is given in Fig.~\ref{fig:plots}, bottom panel.

\begin{figure*}[hbtp!]
  \begin{tabular}{c}
    \psfig{file=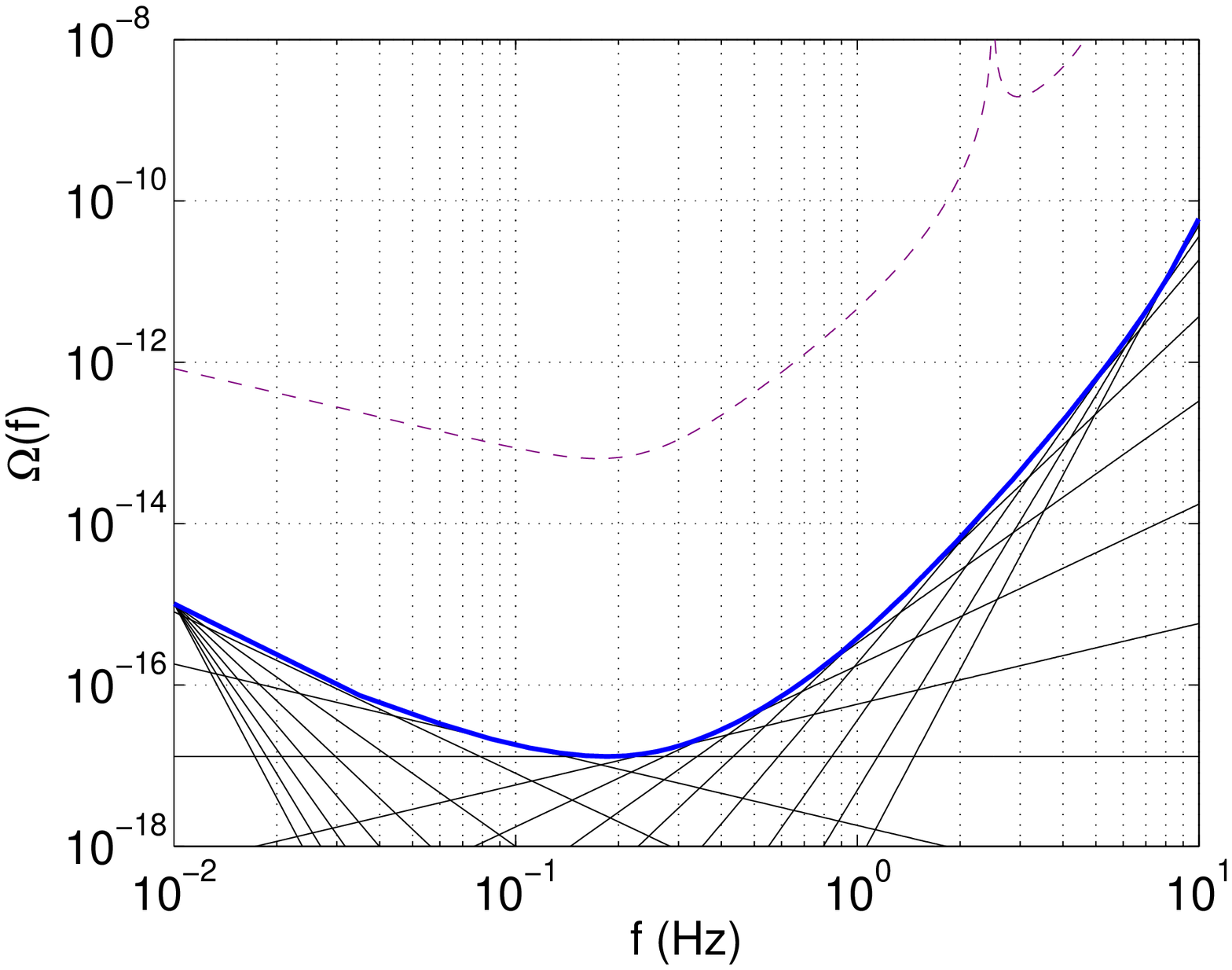, height=2.5in}
    \\
    \psfig{file=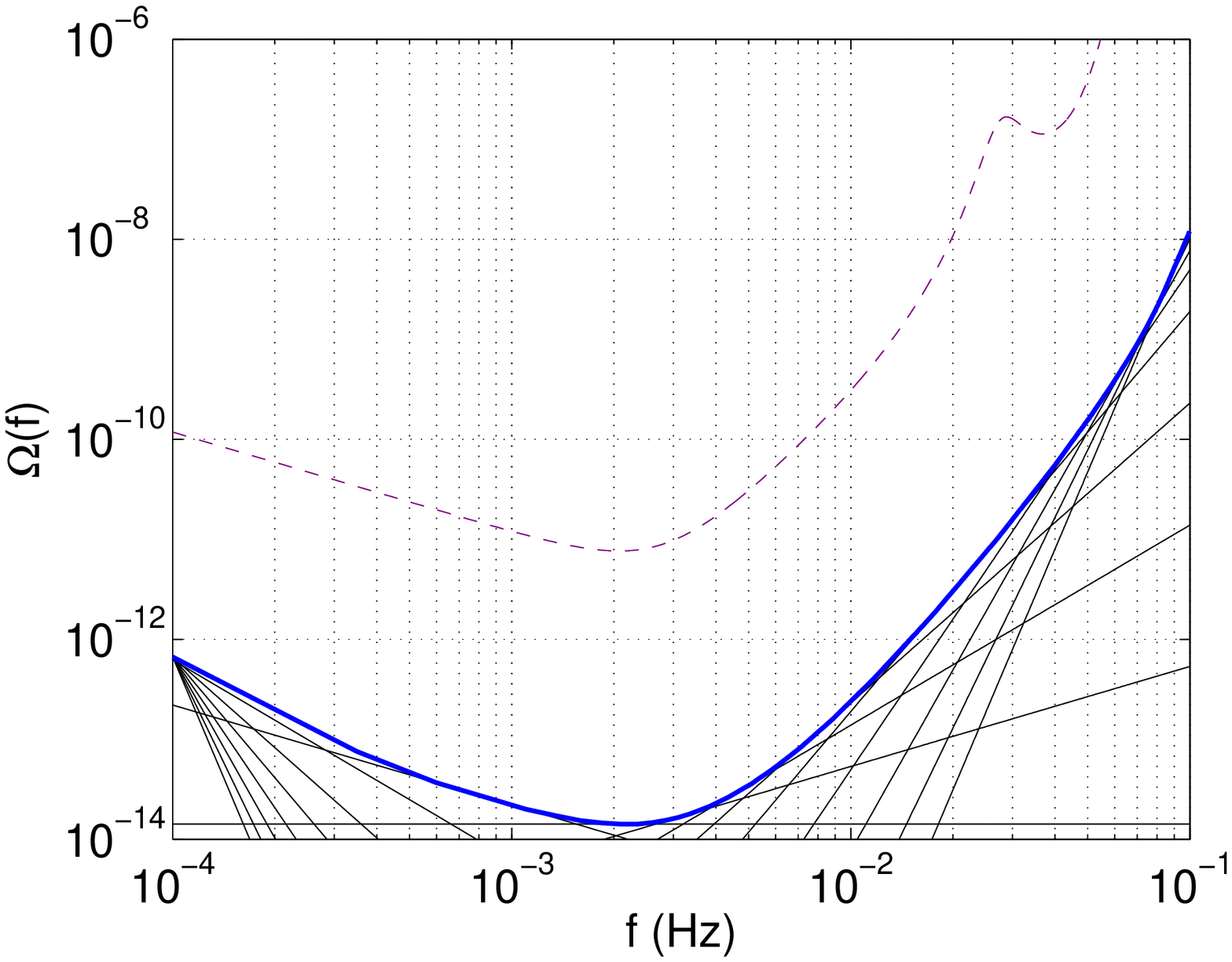, height=2.5in} 
    \\
    \psfig{file=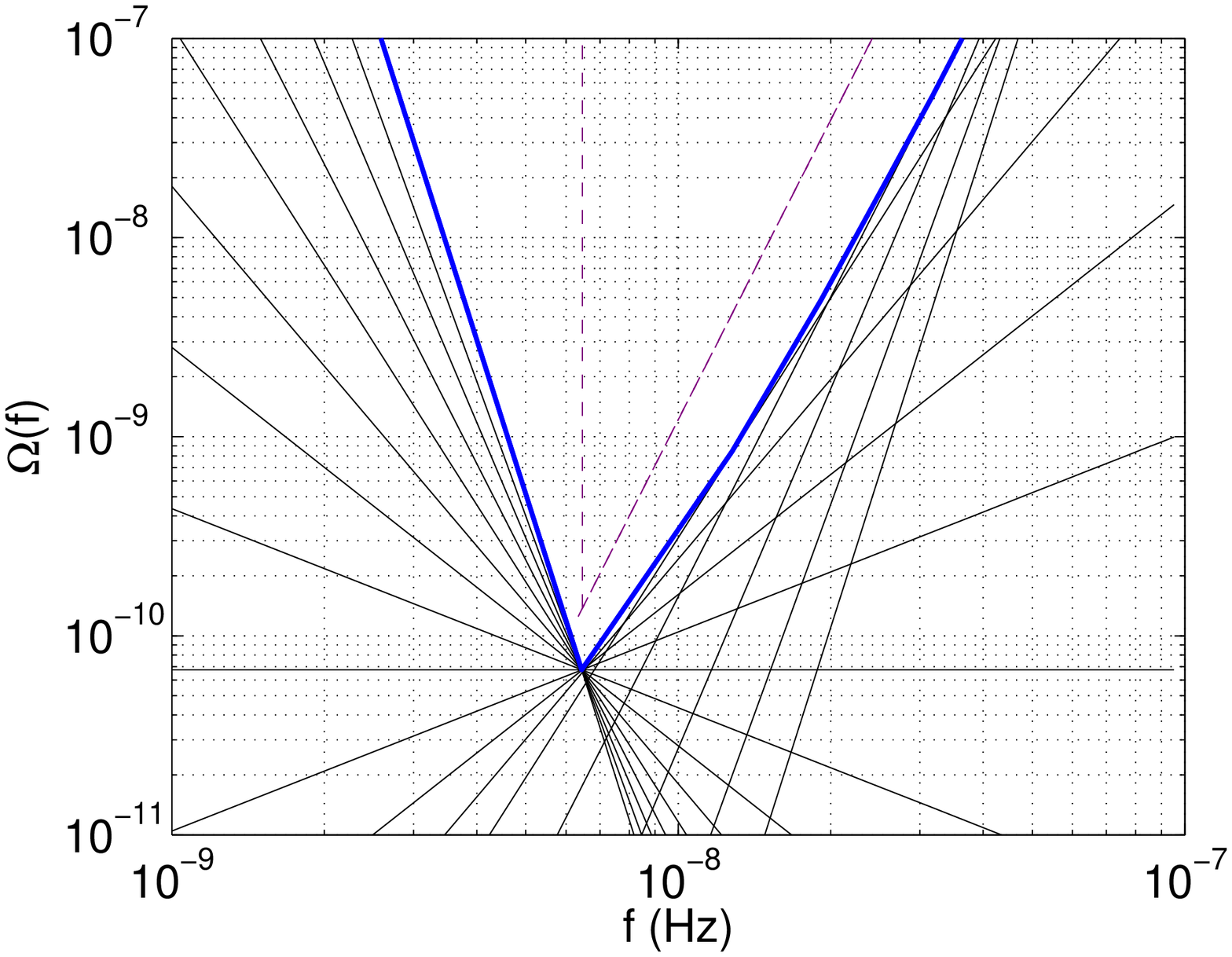, height=2.5in}
  \end{tabular}
  \caption{
    One-sigma, power-law integrated sensitivity curves.
    The dashed purple curves show the effective strain spectral density $S_\text{eff}(f)$ 
    ($S_n(f)$ for LISA, middle panel) converted to fractional energy density units
    (see Eqs.~\ref{e:Seff},~\ref{e:Sn},~\ref{e:Sh}).
    Top panel: BBO assuming $T=\unit[1]{yr}$ of observation.
    The spike at $\approx\unit[2.5]{Hz}$ is due to a zero in the BBO overlap reduction function.
    Middle panel: LISA autocorrelation measurement assuming $T=\unit[1]{yr}$ of observation and
    perfect subtraction of instrumental noise and/or any unwanted astrophysical foreground.
    Bottom panel: A pulsar timing array consisting of 20~pulsars, $\unit[100]{ns}$ timing noise, 
    $T=5~{\rm yr}$ of observation, and a cadence of $\unit[20]{yr^{-1}}$.
  }
  \label{fig:plots}
\end{figure*}

It is interesting to note that the power-law integrated curves for Advanced LIGO and 
BBO are relatively round in shape, whereas the pulsar timing curve is pointy.
(The steep $\Omega(f)\propto f^5$ spectrum can be understood as follows: the transfer function ${\cal R}(f)$ contributes a factor of $f^2$ while the conversion from power to energy density contributes an additional factor of $f^3$.)
This reflects the fact that the sensitivity of pulsar timing measurements is mostly 
determined by a small band of the lowest frequencies in the observing band 
regardless of the spectral shape of the signal.
However, the timing-model fit mentioned above may round out the pointy shape of the PTA sensitivity curve.
We also note that the stochastic background in the PTA band may exhibit variability.
The power-law integrated curves represent the sensitivity to energy density observed at Earth over the course of the measurement.
\medskip

Figure~\ref{fig:landscape2} is a summary the results of this section,
showing the power-law integrated sensitivity 
curves for the different detectors on a single plot spanning a wide range of frequencies.

\begin{figure*}[hbtp!]
    \psfig{file=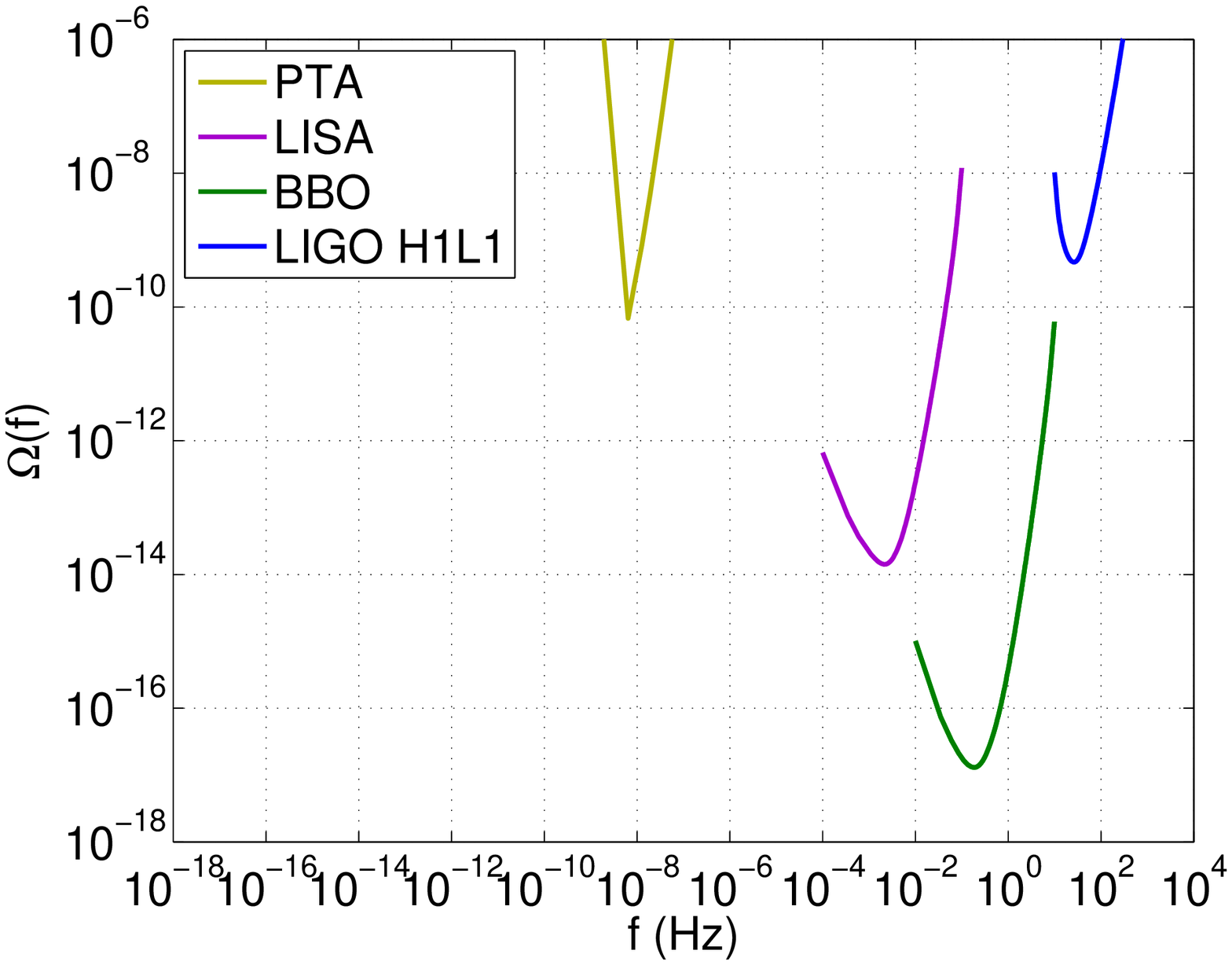, height=4in}
  \caption{
    One-sigma, power-law integrated sensitivity curves
    for the different detectors considered 
    in this paper, plotted on the same graph.
    The Advanced LIGO H1L1, BBO, and pulsar timing sensitivity curves correspond to 
    correlation measurements using two or more detectors.
    The LISA sensitivity curve corresponds to an autocorrelation measurement in a
    single detector assuming perfect subtraction of instrumental noise and/or any 
    unwanted astrophysical foreground.
    }
  \label{fig:landscape2}
\end{figure*}

%%%%%%%%%%%%%%%%%%%%%%%%%%%%%%%%%%%%%%%%%%%%%%%%%%%%%%%%%%%%%%%%%%%%%%%%%%%%%%%
\section{Discussion}
\label{s:discussion}

We have presented a graphical representation of detector sensitivity curves for power-law 
gravitational-wave backgrounds that takes into account the enhancement in sensitivity that 
comes from integrating over frequency in addition to integrating over time.
We applied this method to construct new power-law integrated sensitivity curves for 
cross-correlation searches involving advanced ground-based detectors, BBO, and a network of pulsars 
from a pulsar timing array.
We also constructed a power-law integrated sensitivity curve for an autocorrelation
measurement using LISA.
The new curves paint a more accurate picture of the expected sensitivity of upcoming observations.
The code that we used to produce the new curves is available at 
\url{https://dcc.ligo.org/LIGO-P1300115/public} for public download.
Hopefully, this will allow other researchers to easily construct similar sensitivity curves.
Required inputs are the noise power spectral density 
$P_{nI}(f)$ for each detector in the network and the overlap 
reduction function $\Gamma_{IJ}(f)$ for each detector pair.
Common default files are available for download with the plotting code.

Although the above discussion has focused on comparing predicted strengths of gravitational-wave backgrounds 
to sensitivity curves for current or planned detectors, one can also present measured upper limits for 
power-law backgrounds in a similar way.
That is, instead of plotting the upper limits for $\Omega_\beta$ (for fixed $f_{\rm ref}$) 
as a function of the spectral index $\beta$ as in~\cite{stoch-S5,s5vsr1,paramest}, one can 
plot the envelope of upper-limit power-law curves as a function of frequency.
This would better illustrate the frequency dependence of the upper limits in the observing band of the detectors.

%%%%%%%%%%%%%%%%%%%%%%%%%%%%%%%%%%%%%%%%%%%%%%%%%%%%%%%%%%%%%%%%%%%%%%%%%%%%%%%
\acknowledgments
We thank Vuk Mandic and Nelson Christensen for helpful comments regarding an earlier draft of the paper.
JDR would also like to thank Paul Demorest, Justin Ellis, Shane Larson, Alberto Sesana, and Alberto Vecchio
for discussions related to PTA sensitivity curves.  
ET is a member of the LIGO Laboratory, supported by funding from United States National Science Foundation.
LIGO was constructed by the California Institute of Technology and Massachusetts Institute of Technology 
with funding from the National Science Foundation and operates under cooperative agreement PHY-0757058.
JDR acknowledges support from NSF Awards PHY-1205585, PHY-0855371, and CREST HRD-1242090.

\bibliography{paper}

\end{document}